\documentclass[12pt]{article}
\usepackage{latexsym, epsfig, graphics}
\newcommand{\be}{\begin{equation}}
\newcommand{\ee}{\end{equation}}
\newcommand{\bq}{\begin{eqnarray}}
\newcommand{\eq}{\end{eqnarray}}
\newcommand{\cue}{{\bf q}}
\newcommand{\dbsm}{\sum_{\sigma_{1}<\sigma}\sum_{\sigma<\sigma_{2}}}
\newcommand{\sqsm}{\sqrt{(\sigma-\sigma_{1})(\sigma_{2}-\sigma_{1})
(\sigma_{2}-\sigma)}}
\newcommand{\wss}{W(\sigma_{1},\sigma_{2})}
\newcommand{\ssm}{\frac{\sigma_{2}-\sigma_{1}}{(\sigma_{2}-\sigma)
(\sigma -\sigma_{1})}}
\newcommand{\bfv}{{\bf v}}
\begin{document}
\begin{titlepage}
\today          \hfill 
\begin{center}

\vskip .5in

{\large \bf Field Theory On The World Sheet:Crystal Formation}
\footnote{Notice:* This manuscript has been authored by Korkut Bardakci
under Contract No.DE-AC02-05CH11231 with the U.S. Department of Energy.
The United State Goverment retains and the publisher, by accepting the
article for publication, acknowledges that the United States Goverment
retains a non-exclusive, paid-up, irrecovable, world-wide license to
publish or reproduce the published of this manuscript, or allow others
to do so, for United States Goverment purposes.}
\vskip .50in


\vskip .5in
Korkut Bardakci \footnote{Email: kbardakci@lbl.gov}

{\em Department of Physics\\
University of California at Berkeley\\
   and\\
 Theoretical Physics Group\\
    Lawrence Berkeley National Laboratory\\
      University of California\\
    Berkeley, California 94720}
\end{center}

\vskip .5in

\begin{abstract}
In a previous work, a world sheet field theory which sums planar
$\phi^{3}$ graphs was investigated. In particular, a solitonic
solution of this model was constructed, and quantum fluctuations
around this solution led to a string picture. However, there were
two problems which were not treated satisfactorily: An ultraviolet
divergence and a spurious infrared divergence. Here we present an
improved treatment, which eliminates the ultraviolet divergence
in the normal fashion by mass and coupling constant renormalization.
The infrared problem is taken care of by  choosing a classical
background which forms a one dimensional crystal. The resulting
picture is a hybrid model with both string and underlying field 
theory excitations. Only in the dense graph limit on the world sheet
a full string picture emerges.

\end{abstract}
\end{titlepage}

\newpage
\renewcommand{\thepage}{\arabic{page}}
\setcounter{page}{1}
\noindent{\bf 1. Introduction}
\vskip 9pt

The present work is the continuation of a series of earlier papers [1, 2]
on the same subject.
 It has a lot in common with especially reference [1], but, it
also has two important new features which we think are significant
advances over  [1]. For the convenience of the reader,
 we will first present
a brief discussion of the problem at hand and the results of the earlier
work, and then focus on what is new in the present article.

The idea is to sum the planar graphs of the $\phi^{3}$ field
theory in both $3+1$ and $5+1$ dimensions, starting with the world sheet
picture developed in [3], which in turn was based on the pioneering
work of 't Hooft [4]. This picture, which we briefly review in section 2,
makes use of the mixed light cone parametrization of planar graphs, similar
to the one employed in string theory. In the next section, we describe
the field theory on the world sheet, developed in [5], which reproduces
these graphs. This theory is formulated in terms of a complex scalar
field and a two component fermionic field; a central role is played by the
 field $\rho$ (eq.(4)), a composite of the fermions, which
roughly measures the density of graphs on the world sheet. Just as in
[1,2], here also we are mainly interested in high density graphs, to be
defined more precisely later in terms of $\rho$.
 The basic idea, which motivated some of the
very early work [6, 7], 
  is that a densely covered
world sheet would naturally have a string description. To find such a
string picture  has been a
goal of the present, as well as of the earlier work. 

The world sheet field theory discussed above suffers from two kinds of
divergences: One of them is the standard field theoretic ultraviolent
divergence which we will address later on. The second one is a 
(spurious) infrared divegence due to the choice of the light cone
coordinates. We find our previous treatment of these problems
unsatisfactory, and we readdress them in the present work.
As before, we start by temporarily discretizing the 
$\sigma$ coordinate of the world sheet in steps of length $a$.
 This sort of cutoff has been
extensively used both in field theory [8] and in string theory [9].
Our major goal in this paper is, in addition to eliminating the ultraviolet
divergence by renormalization, to take the limit of zero grid spacing
without encountering any singularity.

Sections 2 and 3 are mostly a review,
but in section 4, we start diverging from the earlier work. An 
important feature of the world sheet field theory is that the existence of
solitonic classical solutions. In [1] and [2], the classical solution was
constructed with the help of the mean field approximation. An unusual
feature of the classical solution is that the corresponding classical
energy is ultraviolet divergent. 
The natural renormalization prescription is  to eliminate this
divergence by means a bare mass counter term. Unfortunately, in [1], the
structure of divergent term did not allow such a cancellation, and
 the interaction vertex had to be modified
 in somewhat ad hoc
fashion in order to achieve such a cancellation. This is an unsatisfactory
 feature of the earlier work which we avoid in the present paper.
This problem can be traced back
 to a somewhat premature application of the mean field
 approxiamation: $\rho$ is originally a kind of spin variable which only
takes on the values 0 and 1, 
 but in the mean field approximation, it
becomes a classical continuous variable in the range $0\leq \rho \leq 1$.
It is this approximate treatment of $\rho$ that is the source of
the trouble. Instead, in section 4, we solve the equations of motion,
treating the scalar field $\phi$ classically, but keeping the fermions
and $\rho$ fully quantum mechanical. We find that mass renormalization
can be  carried out without any ad hoc modifications. This one of the
 new features of the present work that is an improvement over
reference [1].

The main results of section 4 are the two ultraviolet finite expressions
for the classical energy: Eq.(22) in $3+1$ dimensions and eq.(25) in
$5+1$ dimensions. In the latter case, in addition to mass renormalization,
the coupling constant also has to be renormalized. The next step is to
find the field configuration corresponding to the ground state
 that minimizes this energy. Since we cannot do this exactly, 
it is at this point that we introduce the mean field approximation in
section 5, which  was already
used extensively in the previous work. Here, apart from a different starting
point due to renormalization, we also choose a different mean field background
compared to the one chosen in [1] and [2]. The crucial point about this
background, defined by eqs.(27) and (28), is that it vanishes except
on lines equally spaced by a distance $L$ (Fig.4). By letting
the grid space $a$ go to zero while keeping $L$ finite and independent
of $a$, we are able to define a sensible continuum limit. In this limit,
both the classical energy and the quantum corrections about it computed
in section 7, all stay finite. This is to be contrasted with the
$1/a^{2}$ divergence in the classical energy and in some of the spectrum
found in [1] and [2]. We also note that this new background breaks 
translation invariance in $\sigma$; it is therefore natural to
identify it with a one dimensional crystal.

In section 6, we search for the ground state by minimizing the energy
with respect to both $L$ and $\rho_{0}$, the ground state expectation
value of $\rho$. Starting with $3+1$ dimensions, we first vary
with respect to $L$ at fixed $\rho_{0}$ and find a minimum. Next, varying
with respect to  $\rho_{0}$, we find that the energy goes to minus infinity as
$\rho_{0}\rightarrow 1$. Since as $\rho_{0}\rightarrow 1$, 
 $L\rightarrow 0$ (eq.(36)), this is
the limit of densely covered world sheet, which is expected to lead to
string formation. So by lowering its energy, the model is dynamically
driven towards $\rho_{0}\rightarrow 1$. To  investigate this
limit without encountering singular expressions, we introduce a cutoff
on the density of graphs (eq.(40)), which corresponds to an upper bound
on $\rho_{0}$ less than one. It is important to notice that this is
simply a restriction on the choice of graphs; there is no change of the
world sheet dynamics. Later on, in section 8, we discuss the limit 
$\rho_{0}\rightarrow 1$.
In $5+1$ dimensions, things work out differently. The classical
energy vanishes at its minimum, $L$ is determined as a function of 
$\rho_{0}$ by eq.(43) but $\rho_{0}$ itself, at least in this approximation,
 is undetermined.

In the next section, section 7, we compute second order
quantum fluctuations around the classical background. This section is
technically very similar to corresponding material in [1] and [2].
However, there is one important difference: In the previous work, in
limit $a\rightarrow 0$, the energies of many excited states went to infinity.
Here, this limit is smooth, and the excited states all remain at finite
energy. Instead of analyzing the spectrum in full generality, we
we focus on the excitations defined by eq.(49), and determine their
contribution to the action. These states, originally
studied in [1] and [2], are candidates for string excitations.

We investigate the action for the candidate string states in section 8.
This action for $3+1$ dimensions
 is not yet a string action; we show that, only in the
dense graph limit of $\rho_{0}\rightarrow 1$, it tends as a limit
to the light cone string action. A plausible picture of the model for
$\rho_{0}<1$ is the following: The spectrum is a combination of a
heavy sector, consisting of the states of the original field theory,
and lower lying states consisting of string excitations. In the
limit $\rho_{0}\rightarrow 1$, the masses of the heavy states go
to infinity, whereas the string states stay finite. We tentatively
identify a  parameter $\gamma$ (eq.(60)), proportional to
$1-\rho_{0}$, which could serve as an expansion parameter in the
dense graph limit. In section 9, we summarize our conclusions and
 discuss dirctions for future research.

\vskip 9pt
\noindent{\bf 2. The World Sheet Picture}
\vskip 9pt

The planar graphs of $\phi^{3}$ can be represented [4] on a world sheet
parameterized by the light cone coordinates $\tau=x^{+}$ and
$\sigma=p^{+}$ as a collection of horizontal solid lines (Fig.1), where
the n'th line carries a D dimensional transverse momentum $\cue_{n}$.
\begin{figure}[t]
\centerline{\epsfig{file=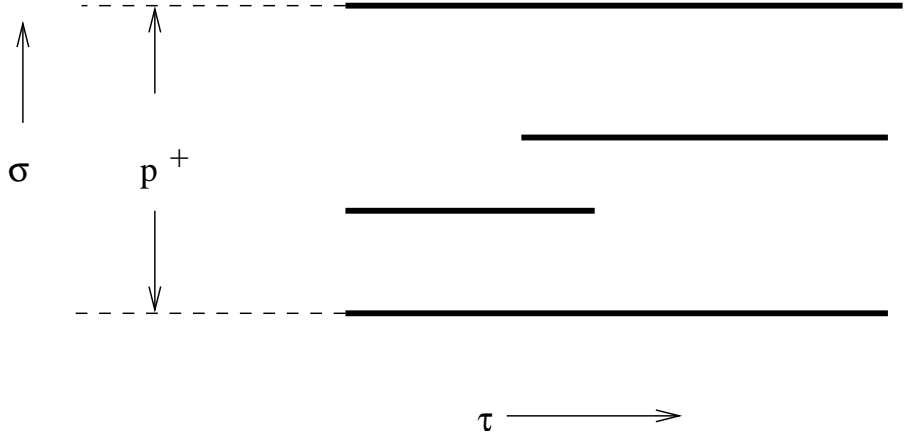, width=12cm}}
\caption{A Typical Graph}
\end{figure}
Two adjacent solid lines labeled by n and n+1 correspond to the light
cone propagator
\be
\Delta({\bf p}_{n})=\frac{\theta(\tau)}{2 p^{+}}\,\exp\left(
-i \tau\, \frac{{\bf p}_{n}^{2}+ m^{2}}{2 p^{+}}\right),
\ee
where ${\bf p}_{n}= \cue_{n}-\cue_{n+1}$ is the momentum flowing through
the propagator. A factor of the coupling constant g is inserted
 at the beginning and at the end of each line, where the interaction
takes place. Ultimately, one has to integrate over all possible
locations and lengths of the solid lines, as well as over the
momenta they carry.

The propagator (1) is singular at $p^{+}=0$. It is well known that 
this is a spurious singularity peculiar to the light cone picture.
To avoid this singularity, and as well as other technical reasons,
it is convenient to temporarily
discretize the $\sigma$ coordinate in steps of length $a$.
A useful way of visualizing the discretized world sheet is
pictured in Fig.2. The boundaries of the propagators are marked by
solid lines as before, and the bulk is filled by dotted lines spaced
at a distance $a$. For the time being, we will keep $a$ finite, and
later, we will show how one can safely take the limit $a\rightarrow 0$.
For convenience, the $\sigma$ is compactified by imposing periodic
boundary conditions at $\sigma=0$ and $\sigma=p^{+}$. In contrast, the
boundary conditions at $\tau=\pm \infty$ are left arbitrary.
\begin{figure}[t]
\centerline{\epsfig{file=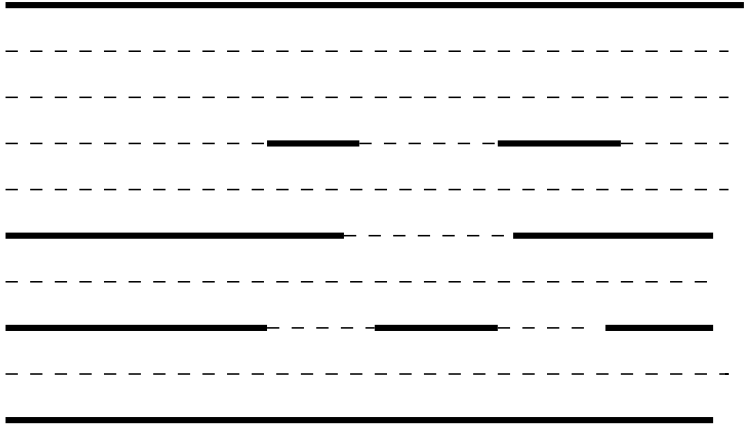, width=10cm}}
\caption{Solid And Dotted Lines}
\end{figure}

\vskip 9pt

\noindent{\bf 3. The World Sheet Field Theory}

\vskip 9pt

It was shown in [5] that the light cone graphs described above are
reproduced by a world sheet field theory, which we now briefly review.
We introduce the complex scalar field $\phi(\sigma,\tau,\cue)$ and
its conjugate $\phi^{\dagger}$, which at time $\tau$
 annihilate (create) a solid line with coordinate $\sigma$ carrying
momentum $\cue$. They satisfy the usual commutation relations
\be
[\phi(\sigma,\tau,\cue),\phi^{\dagger}(\sigma',\tau,\cue')]=
\delta_{\sigma,\sigma'}\,\delta(\cue-\cue').
\ee
The vacuum, annihilated by the $\phi$'s, represents the empty world sheet.

In addition, we introduce a two component fermion field $\psi_{i}(
\sigma,\tau)$, $i=1,2$, and its adjoint $\bar{\psi}_{i}$, which
satisfy the standard anticommutation relations. The fermion with
$i=1$ is associated with the dotted lines and $i=2$ with the solid
lines. The fermions are needed to avoid unwanted configurations
on the world sheet. For example, multiple solid lines generated by
the repeated application of $\phi^{\dagger}$ at the same $\sigma$
would lead to overcounting of the graphs. These redundant states can
be eliminated by imposing the constraint
\be
\int d\cue\, \phi^{\dagger}(\sigma,\tau,\cue)\phi(\sigma,\tau,\cue)
=\rho(\sigma,\tau),
\ee
where
\be
\rho=\bar{\psi}_{2}\psi_{2},
\ee
which is equal to one on solid lines and zero on dotted lines. This
constraint ensures that there is at most one solid line at each
site.

Fermions are also needed to avoid another set of unwanted configurations.
Propagators are assigned only to adjacent solid lines and not to
non-adjacent ones. To enforce this condition, it is convanient to
define, 
\be
\mathcal{E}(\sigma_{i},\sigma_{j})=\prod_{k=i+1}^{k=j-1}\left(
1-\rho(\sigma_{k})\right),
\ee
for $\sigma_{j}>\sigma_{i}$, and zero for $\sigma_{j}<\sigma_{i}$.
The crucial property of this function is that it acts as a projection: 
It is equal to one when the two lines at $\sigma_{i}$ and $\sigma_{j}$
are seperated only by the dotted lines; otherwise, it is zero. With the
help of $\mathcal{E}$, the free Hamiltonian can be written as
\bq
H_{0}&=&\frac{1}{2}
\sum_{\sigma,\sigma'}\int d\cue \int d\cue'\,\frac{\mathcal
{E}(\sigma,\sigma')}{\sigma'-\sigma} \left((\cue-\cue')^{2}+ m^{2}
\right)\nonumber\\
&\times& \phi^{\dagger}(\sigma,\cue) \phi(\sigma,\cue)
 \phi^{\dagger}(\sigma',\cue') \phi(\sigma',\cue')\nonumber\\
&+&\sum_{\sigma} \lambda(\sigma)\left(\int d\cue\,
 \phi^{\dagger}(\sigma,\cue) \phi(\sigma,\cue) -\rho(\sigma)\right),
\eq
where $\lambda$ is a lagrange multiplier enforcing the constraint (3).
The evolution operator $\exp(-i \tau H_{0})$, applied to states,
generates a collection of free propagators, without, however, the
prefactor $1/(2 p^{+})$.

Using the constraint (3), the free hamiltonian can be written in a
form more convenient for later application:
\bq
H_{0}&=&\frac{1}{2}\sum_{\sigma,\sigma'}G(\sigma,\sigma')\Bigg(
\frac{1}{2} m^{2}\,\rho(\sigma) \rho(\sigma') + \rho(\sigma')\,
\int d\cue\,\cue^{2}\, \phi^{\dagger}(\sigma,\cue) \phi(\sigma,\cue)
\nonumber\\
&-&\int d\cue \int d\cue'\,(\cue\cdot \cue')\,
\phi^{\dagger}(\sigma,\cue) \phi(\sigma,\cue)
\phi^{\dagger}(\sigma',\cue') \phi(\sigma',\cue')\Bigg)\nonumber\\
&+&\lambda(\sigma)\left(\int d\cue\,
 \phi^{\dagger}(\sigma,\cue) \phi(\sigma,\cue) -\rho(\sigma)\right),
\eq
where we have defined
\be
G(\sigma,\sigma')=\frac{\mathcal{E}(\sigma,\sigma')+
\mathcal{E}(\sigma',\sigma)}{|\sigma-\sigma'|}.
\ee

Next, we introduce the interaction term. Two kinds of interaction
vertices, corresponding to $\phi^{\dagger}$ creating a solid line
or $\phi$ destroying a solid line, are pictured in Fig.3.
\begin{figure}[t]
\centerline{\epsfig{file=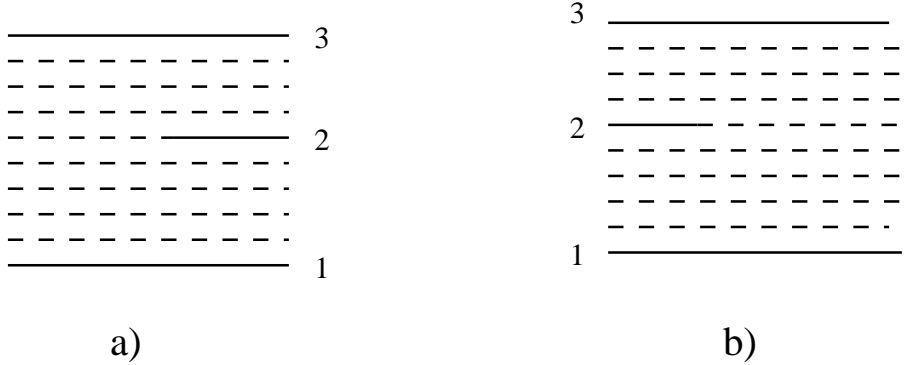, width=12cm}}
\caption{The Two $\phi^{3}$ Vertices}
\end{figure}
 We also
have to take care of the prefactor $1/(2 p^{+})$ in (1) by attaching it
to the vertices. Here, as in [5], we choose a symmeteric distribution of 
this factor, by attaching a factor of
\be
V=\frac{1}{\sqrt{8\, p_{12}^{+}\, p_{23}^{+}\, p_{13}^{+}}}=
\frac{1}{\sqrt{8\,(\sigma_{2}-\sigma_{1})(\sigma_{3}-\sigma_{2}) 
(\sigma_{3}-\sigma_{1})}}
\ee
to each vertex. Different ways of splitting the prefactor $1/(2 p^{+})$   
result in non-symmetric vertices; here we choose the standard
symmetric form. The interaction term in the hamiltonian can
now be written as
\be
H_{I}= g\,\sum_{\sigma}\int d\cue\,\left(\mathcal{V}(\sigma)\,
\rho_{+}(\sigma)\, \phi(\sigma,\cue)+
\rho_{-}(\sigma)\,\mathcal{V}(\sigma)\,
\phi^{\dagger}(\sigma,\cue)\right),
\ee
where $g$ is the coupling constant. $\rho_{\pm}$ are given by
$$
\rho_{+}=\bar{\psi}_{1} \psi_{2},\,\,\,\rho_{-}=\bar{\psi}_{2}
\psi_{1},
$$
and
\be
\mathcal{V}(\sigma)=\dbsm \frac{W(\sigma_{1},\sigma,\sigma_{2})}
{\sqsm},
\ee
where,
\be
\wss=\rho(\sigma_{1})\, \mathcal{E}(\sigma_{1},\sigma_{2})\,
 \rho(\sigma_{2}).
\ee

Here is a brief explanation of the origin of various terms in $H_{I}$:
The factors of $\rho_{\pm}$ are there to pair a solid line
with an $i=2$ fermion and a dotted line with an $i=1$ fermion. The
factor of $\mathcal{V}$ ensures that the pair of solid lines 12 and 23
in Fig.3
are seperated by only dotted lines, without any intervening solid lines.
Apart from an overall factor, the vertex defined above is very similar
 to the bosonic string interaction vertex in the light cone
picture. Taking advantage of the properties of $\mathcal{E}$ discussed
following eq.(5), we have written an explicit representation of 
this overlap vertex.

Finally, the total hamiltonian is given by
$$
H=H_{0}+H_{I}
$$
and the corresponding action by
\be
S=\int d\tau\left(\sum_{\sigma}\left(i \bar{\psi} \partial_{\tau}
\psi + i\int d\cue\,\phi^{\dagger} \partial_{\tau} \phi \right)
- H(\tau)\right).
\ee

\vskip 9pt

\noindent{\bf 4. The Semi Classical Solution And Renormalization}

\vskip 9pt

In this section, our goal is to search for classical
solutions to the equations of motion that follow from the action (7).
The idea is to eliminate terms linear in $\phi$ and $\phi^{\dagger}$
in $H_{I}$ by setting
\be
\phi=\phi_{0}+ \phi_{1},\,\,\,\phi^{\dagger}=\phi^{\dagger}_{0}
+\phi_{1}^{\dagger}.
\ee
Here, $\phi_{0}$ is the  field fixed by the equations of motion,
and $\phi_{1}$ represents the  fluctuations around the classical
background. There remains the question of what to do about the
fermions. In the previous work [1,2], fermions were bosonized, and the
fermionic bilinears $\rho,\,\rho_{\pm}$ were the treated as fixed
 classical background fields. In contrast, here, for the time being,
 the fermions, as well as the field $\lambda$ will be treated exactly;
 in particular, we wish to preserve the relations
\be
\rho^{2}(\sigma)=\rho(\sigma),\,\,\rho_{+}(\sigma)\rho_{-}(\sigma)= 
1-\rho(\sigma),\,\,\rho_{-}(\sigma)\rho_{+}(\sigma)=\rho(\sigma),
\ee
which follow from $\rho$ being a discrete variable, taking on only the values
zero and one.
 As we shall shortly see,
they are needed to show that the self mass divergence
can be absorbed into the mass term already present in the action, without
 need for ad hoc counter terms. The field $\phi_{0}$ is a kind of
hybrid: The equations of motion for $\phi$ are used, but the fermions 
 are still fully quantum mechanical. This is why we use
the term semi classical.

We choose $\phi_{0}$ so that it depends only on $\cue^{2}$ (rotation
 invariance). As a result, the  term that has the factor
$\cue\cdot \cue'$ on the right hand side of eq.(7)
does not contribute, and the equation of motion for $\phi_{0}$
reduces to
\be
\left(\frac{1}{2} G(\sigma,\sigma')\,\rho(\sigma')\,\cue^{2}
+\lambda(\sigma)\right)\,\phi_{0}(\sigma,\cue)+ g\,\rho_{-}(\sigma)
\,\mathcal{V}(\sigma)=0,
\ee
 with a conjugate equation for $\phi_{0}^{\dagger}$. The solution
can be written as
\bq
\phi_{0}(\sigma,\cue)&=& -g\,\frac{\rho_{-}(\sigma)
\,\mathcal{V}(\sigma)}{\lambda(\sigma)+
\frac{1}{2} G(\sigma,\sigma')\,\rho(\sigma')\,\cue^{2}}\nonumber\\
 &=& -g \dbsm\frac{\rho_{-}(\sigma)\,\wss}{\left(\lambda(\sigma)+
\frac{1}{2} \cue^{2}\left(\ssm\right)\right) \sqsm}.\nonumber\\
& &
\eq

To derive the second line in this equation, we expand the denominator
in powers of $\frac{1}{2} G(\sigma,\sigma')\,\rho(\sigma')\,\cue^{2}$
and repeatedly use the identity
\bq
G(\sigma,\sigma')\, \rho(\sigma')\,\rho_{-}(\sigma)\,\wss &=&\nonumber\\
\left(\delta_{\sigma',\sigma_{2}}\,\frac{1}{\sigma_{2}- \sigma}+
\delta_{\sigma',\sigma_{1}}\,\frac{1}{\sigma -\sigma_{1}}\right)
&\rho_{-}(\sigma)&\wss,
\eq
which can easily be derived by expressing $G$ and $W$ in terms of
the $\rho$'s (eqs. (8,12)), and making use of the relations (15). We should
also mention that since $\rho_{\pm}$ do not commute with $\rho$, the
ordering of these factors matters, and the factors $\rho_{-}$ and $W$
 in eq.(17) and (18) are ordered correctly.

We can now compute the classical hamiltonian $H_{c}$ by letting
$\phi\rightarrow \phi_{0}$ in $H$ (eqs.(7,10)). There is,
however, an ultraviolet divergence, resulting from the integration
over $\cue$, which has to be addressed.
 So far, the transverse
dimension $D$ has been arbitrary, but now we have to make a choice.
We specialize to the case $D=2$, where the only ultraviolet divergence
is a logarithmic divergence in the self mass.
 The result is
\bq
H_{c}&=&- 2\pi\,g^{2}\,\sum_{\sigma}\dbsm \frac{\wss}{(\sigma_{2}
-\sigma_{1})^{2}}\,\ln\left(\frac{\Lambda^{2}}{\lambda(\sigma)}
\ssm\right)\nonumber\\
&-& \sum_{\sigma} \lambda(\sigma)\,\rho(\sigma),
\eq
where $\Lambda$ is an ultraviolet cutoff needed because of the mass
divergence. In deriving this result, one needs the identity
\be
\wss \rho_{+}(\sigma) \rho_{-}(\sigma) W(\sigma'_{1},\sigma'_{2})
=\delta_{\sigma_{1},\sigma'_{1}}\,\delta_{\sigma_{2},\sigma'_{2}}   
\,\wss,
\ee
which follows from the definition of $\wss$ (eq.(12)) and the identities
(15). One can also understand it geometrically from the overlap
properties of the vertices in Fig.3.  Apart
from an overall factor, these are structurally the same as the
corresponding string vertices, and in particular, they satisfy the
same overlap relations.

The classical hamiltonian can be renormalized by replacing the cutoff
$\Lambda$ by a an arbitrary finite mass $\mu$. This amounts to introducing
a counter term
\be
2\pi g^{2}\,\sum_{\sigma} \dbsm \frac{\wss}{(\sigma_{2}-\sigma_{1})^{2}}
\,\ln\left(\Lambda^{2}/\mu^{2}\right)= \frac{2 \pi\,g^{2}}{a} \dbsm
\frac{\wss}{|\sigma_{1}-\sigma_{2}|}\,\ln\left(\Lambda^{2}/\mu^{2}\right).
\ee
 Noticing that the term to be summed over is $\sigma$
independent, the sum over this variable was done explicitly.
Now, since, from their definition,
$$
\rho(\sigma)\,G(\sigma,\sigma')\,\rho(\sigma')=
\frac{W(\sigma,\sigma')+W(\sigma',\sigma)}{|\sigma-\sigma'|},
$$
 the above counter term is simply proportional
to the $m^{2}$ term on the right hand side of (7). It can therefore be 
identified with a  cutoff dependent part of the
mass term. In fact, after eliminating the cutoff dependent part,
the remaining finite portion of the mass term can be completely
 absorbed into the definition of $\mu^{2}$, and
 from now on, we shall
assume that this has been done.
The renormalized classical hamiltonian is then given by
\bq
H_{c}^{r}&=&- 2\pi\,g^{2}\,\sum_{\sigma}\dbsm \frac{\wss}{(\sigma_{2}
-\sigma_{1})^{2}}\,\ln\left(\frac{\mu^{2}}{\lambda(\sigma)}
\ssm\right)\nonumber\\
&-& \sum_{\sigma} \lambda(\sigma)\,\rho(\sigma),
\eq

We would like to emphasize that so far no approximation has been
made, and therefore the above equation for the classical part of the
hamiltonian is exact. This why the mass renormalization can be
carried out without introducing ad hoc terms not present in the
original action. Of course, $H_{c}$ is not the whole story; terms 
that depend on
$\phi_{1}$, as well as terms involving derivatives with respect to
$\tau$ are not present in the classical hamiltonian. In the following
sections, we will carry out an expansion to second in the fluctuations around
the classical solutions, and show that the terms we compute
 are all ultraviolet finite.

Next we consider $D=4$, corresponding to $\phi^{3}$ in six dimensions.
The self mass is now quadratically divergent, but this divergence can 
be eliminated by a mass counter term exactly as in the case $D=2$.
There remains, however, a residual logarithmic divergence:
\bq
H_{c}&=& -4 \pi^{2}\,g_{0}^{2}\,\sum_{\sigma}\dbsm \lambda(\sigma)
\,\wss \,\frac{(\sigma_{2}-\sigma) (\sigma -\sigma_{1})}
{(\sigma_{2} -\sigma_{1})^{3}}\nonumber\\
&\times& \ln\left(\frac{\lambda(\sigma)}{\Lambda^{2}} \ssm\right)
-\sum_{\sigma} \lambda(\sigma)\,\rho(\sigma).
\eq
This divergence can be eliminated by renormalizing the bare coupling
constant $g_{0}$ by setting
\be
g_{0}^{2}=\frac{g_{r}^{2}}{\ln\left(\Lambda^{2}/\mu^{2}\right)},
\ee
where $g_{r}$ is the renormalized coupling constant and $\mu$ an
arbitrary mass parameter. We recall that $\phi^{3}$ is asymptotically free
in 6 space-time dimensions, and the above relation between the bare and
renormalized couplings is the well known lowest order result.
 In the limit $\Lambda\rightarrow \infty$,
the renormalized $H_{c}$ is given by
\be
H_{c}^{r}= 4 \pi^{2} g_{r}^{2} \sum_{\sigma} \dbsm \lambda(\sigma)
\left(\wss \,\frac{(\sigma_{2}-\sigma) (\sigma -\sigma_{1})}
{(\sigma_{2} -\sigma_{1})^{3}} -\rho(\sigma)\right).
\ee

\vskip 9pt

\noindent{\bf 5. The Meanfield Approximation}

\vskip 9pt

As we have already pointed out, the expressions for $H_{c}^{r}$ for
$D=2$ and $D=4$ (eqs.(22, 25)) are exact. 
The fermionic fields $\psi$ and $\bar{\psi}$ and the lagrange multiplier
 $\lambda$ are still
fully quantum mechanical, to be integrated over in the functional
integral. Clearly, it is not possible to do the functional integrals
indicated above exactly. It is at this point that we finally have
to introduce some sort of approximation. The scheme we choose is
the mean field method, already extensively used in the previous
work on this subject [1,2]. The reason for postponing the introduction
of this approximation is connected with renormalization: We were able
to absorb the self mass divergence into the mass term already present in
the action by making use of the overlap relations (20). These relations in turn
 followed from the identities (15), which are  violated in
the mean field approximation. By  avoiding  any
approximation before deriving
 the finite renormalized expressions  for $H^{r}_{c}$,
we are able to bypass this problem.

The mean field approximation amounts to expanding the fields $\rho$, $\lambda$
and $\phi$
about their classical expectation values $\rho_{c}$ and $\lambda_{c}$:
\be
\rho(\sigma)=\rho_{c}(\sigma)+\rho_{1}(\sigma),\,\,\,
\lambda(\sigma)=\lambda_{c}(\sigma)+\lambda_{1}(\sigma)\,\,\,
\phi(\sigma)=\phi_{c}(\sigma)+\phi_{1}(\sigma),
\ee
where $0\leq \rho_{c} \leq 1$.
In the earlier work,  $\rho_{0}(\sigma)$ and $\lambda_{0}(\sigma)$
were taken to be constants independent of $\sigma$, and also, of course,
of $\tau$. Here, we make a different choice for these classical
background fields. We set
\be
\rho_{c}(\sigma)= \rho_{0},\,\,\,\lambda_{c}(\sigma)=\lambda_{0},
\ee
at $\sigma=\sigma_{0}+n L$, and
\be
\rho_{c}(\sigma)=0,\,\,\,\lambda_{c}(\sigma)=0,
\ee
for $\sigma\neq \sigma_{0}+n L$. Here n runs over all integers, and
$\sigma_{0}$ is arbitrary. We have also effectively let
$p^{+}\rightarrow \infty$, and we will discuss this later on.
 This structure is pictured
in Fig.4: $\rho_{c}$ and $\lambda_{c}$ are  constants $\rho_{0}$
and $\lambda_{0}$ on what we call hybrid lines,
  seperated by intervals of distance $L$,
and they vanish elsewhere.
\begin{figure}[t]
\centerline{\epsfig{file=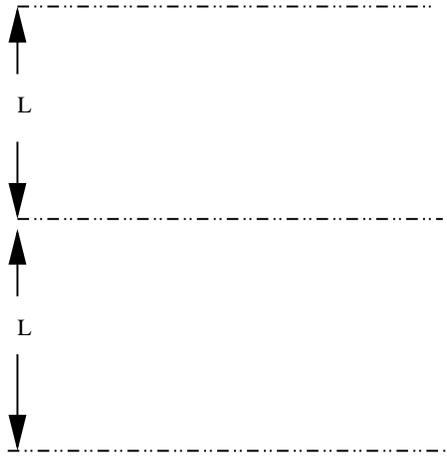, width=6cm}}
\caption{Equally Spaced Hybrid Lines}
\end{figure}
 Hybrid lines, to be defined
more precisely in section 5, are superpositions of solid
and dotted lines.
 Since $\sigma$ is discretized in units of $a$,
L is an integer multiple of $a$. The important point is that as we
 eventually let $a\rightarrow 0$, $L$ will be kept fixed and finite.

It remains to specify $\phi_{c}$.
We  take
\be
\phi_{c}= 0
\ee 
at  $\sigma\neq \sigma_{0}+ n L$, and at $\sigma=\sigma_{0}+n L$,
 we simply set $\rho=\rho_{0}$
and $\lambda=\lambda_{0}$ in eq.(17) for $\phi_{0}$. The only remaining
question is what to do about $\rho_{\pm}$. Actualy, since $H_{c}$
depends only on $\rho$, we do not
 really need to know $\rho_{\pm}$ individually. However,
 for the sake of completeness, we note that, in the
classical approximation,
\be
\rho_{+}=\rho_{-}= \sqrt{\rho -\rho^{2}}.
\ee
 This follows both from the bosonization of the fermions [1,2], and
also from the mean field approximation we will discuss shortly. We
finally note that
 the classical fields are non-trivial
only on the hybrid lines; they are zero elsewhere. This will be important
for having a finite $a\rightarrow 0$ limit.

 The mean field approximation $\rho(\sigma)\rightarrow \rho_{c}(\sigma)$
 amounts to replacing $\rho$ by its expectation value
\be
\rho_{c}(\sigma)= \langle \sigma|\rho(\sigma)|\sigma \rangle,
\ee
where the fermionic state $|\sigma\rangle$ is defined by
\be
|\sigma\rangle = \left(\sqrt{\rho_{0}}\,\,\bar{\psi}_{2}(\sigma)+
\sqrt{1 -\rho_{0}}\,\,\bar{\psi}_{1}(\sigma)\right) |0\rangle,
\ee
for $\sigma= \sigma_{0}+n L$, and,
\be
|\sigma\rangle = |0\rangle, 
\ee
for $\sigma \neq \sigma_{0}+ n L$. What was represented by a
 hybrid line in Fig.4 is
concretely the state $|\sigma\rangle$, a linear superposition of solid  
and dotted lines. This state can also be thought of as a
variational ansatz [5] for minimizing the classical energy with respect
to the parameter $\rho_{0}$. The novelty of this approach compared to the
earlier work is that in addition to $\rho_{0}$, we will also treat
$L$ as a variational parameter. Whereas in [1,2], $L$ is
in effect set equal to the lattice spacing $a$, here it is freed
from that restriction and allowed to vary in order to minimize
the classical energy.

It is clearly advantageous to have as many free parameters as possible
in a variational calculation, so the freeing of $L$ from the restriction
$L=a$ is a welcome development. This new flexibility  has another big
advantage: So long as $L$ is kept fixed, the limit $a\rightarrow 0$
is smooth (non-singular). This point will be discussed further in the
following sections, but it can already be gleaned from eq.(6). If one
sets $L=a$, two hybrid lines are allowed to be at a distance $a$ apart.
Since the hybrid lines are part of the time solid lines, this means that
the denominator $\sigma' -\sigma$ of the first term in eq.(6) for
$H_{0}$ can be equal to $a$, becoming singular as  $a\rightarrow 0$.
On the other hand, if the hybrid lines are a distance $L$ apart, the
 minimum value of this denominator is $L$, which remains finite as
$a\rightarrow 0$. In fact, in the next section, we will show that
the value of $L$ that minimizes $H_{c}$ depends only on $\rho_{0}$
and it is independent of $a$.

How do we know that the classical background we have chosen is the
right one? We have chosen it because it was the simplest regular
configuration we could think of which was non-singular in the limit
$a\rightarrow 0$. It then clearly has lower energy the the 
background chosen in reference [1], where the classical energy
becomes infinite in the limit $a\rightarrow 0$.
 Of course, one could not exclude the possibility
of more complicated backgrounds with even lower classical energy.

\vskip 9pt
 \noindent{\bf 6. The Ground State Of The Model}

\vskip 9pt

We start with the model at $D=2$ (3+1 space-time dimensions).
To find the ground state, $H^{r}_{c}$ of
eq.(22) has to be minimized  with respect to the chosen classical
background. We note that in the triple sum, the
$\sigma$'s are restricted to the hybrid lines at $\sigma=\sigma_{0}+ n L$,
 and at
these locations, $\rho$ and $\lambda$ are given by eqs.(27) and (28).
This means that, for example, we evaluate $\wss$  by setting
$\rho(\sigma_{1})=\rho(\sigma_{2})=\rho_{0}$ 
 and  by assigning a factor of $1-\rho_{0}$ to each hybrid line
in between $\sigma_{1}$ and $\sigma_{2}$ in eq.(12).
As a result
\be
H_{c}^{r}=-\frac{2 \pi\,g^{2}\,\rho_{0}^{2}\,p^{+}}{L^{3}}\,
\sum_{n_{1}=1}^{\infty} \sum_{n_{2}=1}^{\infty}\frac{(1-\rho_{0})^
{n_{1}+n_{2} -1}}{(n_{1}+n_{2})^{2}}\,\ln\left(\frac{\mu^{2}}{
\lambda_{0} L}\,\frac{n_{1}+n_{2}}{n_{1} n_{2}}\right)
-\frac{\lambda_{0} \,\rho_{0}\,p^{+}}{L}.
\ee

Strictly speaking, the above sums over the n's, instead of going all the
way to infinity, should have an upper cutoff of the order of
$p^{+}/L$, since $\sigma$ is restricted by $0\leq \sigma \leq p^{+}$.
However, this only makes a difference for small values of $\rho_{0}$,
and in what follows, we will be interested only the values of $\rho_{0}$
near one, for which it is a good approximation to let the upper limit
go to infinity. This equivalent to decompactifying the world sheet
by letting $p^{+}\rightarrow \infty$, as was done earlier.

Next, we minimize $H_{c}^{r}$ with respect to $\lambda_{0}$ and $L$
at fixed $\rho_{0}$ by setting
\be
\frac{\partial H_{c}^{r}}{\partial\lambda_{0}}=0,\,\,\,
\frac{\partial H_{c}^{r}}{\partial L}=0.
\ee
The solution to these equations is
\bq
\lambda_{0}&=& \frac{2 \pi\,g^{2}\,\rho_{0}\,(1 -\rho_{0})}{L^{2}}\,
f_{1}(\rho_{0}),\nonumber\\
L&=&\frac{g^{2}\,\rho_{0} (1-\rho_{0})\,f_{1}(\rho_{0})}{\mu^{2}}\,
\exp\left(\frac{1}{3}- \frac{f_{2}(\rho_{0})}{f_{1}(\rho_{0})}\right),
\eq
where,
\bq
f_{1}(\rho_{0})&=&\sum_{n=0}^{\infty} \frac{n+1}{(n+2)^{2}}\,(1-\rho_{0})^
{n},\nonumber\\
f_{2}(\rho_{0})&=& \sum_{n_{1}=1}^{\infty} \sum_{n_{2}=1}^{\infty}
\frac{(1-\rho_{0})^{n_{1}+n_{2} -2}}{(n_{1}+n_{2})^{2}}\,\ln\left(
\frac{n_{1}+n_{2}}{n_{1}\,n_{2}}\right),
\eq
and
the classical hamiltonian (energy) at these values of $L$ and $\lambda_{0}$
and at some fixed value of $\rho_{0}$ reduces to
\be
H_{c}^{r}=-\frac{2 \pi\,\mu^{6}\,p^{+}}{3\,g^{4}\,\rho_{0}\,
(1-\rho_{0})^{2}\,f_{1}^{2}(\rho_{0})}\,\exp\left(3\,\frac{f_{2}(\rho_{0})}
{f_{1}(\rho_{0})}\, -1\right).
\ee
We note that\\
a) The dependence on  $a$, the  spacing of the grid in $\sigma$,
 has completely disappeared. Therefore, the limit
 $a\rightarrow 0$ is trivial.\\
b) A new discrete structure has emerged: These are the hybrid lines
spaced at intervals of length $L$. Notice, however, that the world sheet
remains smooth, and the hybrid lines are effectively zero branes
placed at regular intervals in the bulk. Their spacing $L$ is fixed,
but the location $\sigma_{0}$ (eq.(28)) is arbitrary. The energy is
independent of $\sigma_{0}$ because of translation invariance in 
 $\sigma$. By fixing
$\sigma_{0}$, this invariance is (spontaneously) broken. We are therefore
witnessing the formation of a one dimensional crystal. One can then
identify  the fluctuations about the classical solution with the
vibrations of this crystal, among them will be a zero mode corresponding
to the broken translation invariance. The inclusion of this mode will
presumably restore translation invariance. We will have more to say
about this in the next section.\\
c) The classical energy (38) is negative. Since $H_{0}$ of eq.(6) is
positive semi-definite, it is the interaction that is responsible
for changing the sign of the energy. Both the sign of the energy
and the existence of an optimal spacing $L$ can be understood as
follows: In the expression  for  $H_{0}$, the first term acts as a
repulsive potential between to adjacent hybrid lines and pushes them
apart. But there is an entropic attractive force which balances
this repulsion and leads to a stable configuration. The origin of the
entropic force has to do with the counting of configurations. 
$H_{c}^{r}$ calculated above is really the free energy
$$
F= E- T S,
$$
 which takes into account the entropy arising from the counting of 
configurations. A hybrid line involves  transitions between solid and
dotted lines (Fig.(5)), and as such represents the superposition
of a multitude of configurations, giving rise to increased entropy.
The entropic force and the negative sign of the free energy comes
from the contribution of the hybrid lines to the entropy.
 The entropic term in the free energy favors the increase
in the number of hybrid lines, hence leading to their close spacing.
Balancing this is the repulsive term in  $H_{0}$, which is
 trying to keep the hybrid lines apart.

So far, $\rho_{0}$ has been fixed in the interval $0\leq \rho_{0}
\leq 1$. We should now consider minimizing $H_{c}^{r}$ with respect
to $\rho_{0}$. It turns out that  $H_{c}^{r}$ is steadily decreasing
function which goes from zero at $\rho_{0}=0$ 
 to $-\infty$ as $\rho_{0}\rightarrow 1$:
\be
 H_{c}^{r}\rightarrow -\frac{16 \pi\,p^{+}\,\mu^{6}}{3\,g^{4}\,e
(1-\rho_{0})^{2}}.
\ee
This clearly a singular limit which we will investigate in more
detail later on, meanwhile, we will introduce a cutoff on the average 
number of solid lines on the world sheet by setting
\be
\sum_{\sigma} \rho(\sigma)\rightarrow\frac{\rho_{0}\,p^{+}}{L}\leq \kappa,
\ee
where $\kappa$ is a fixed constant. Since $\rho_{0}\rightarrow 1$ corresponds
to $\kappa\rightarrow \infty$, a finite $\kappa$ corresponds to an
 upper bound on $\rho_{0}$ less than
unity, and it is at this value of $\rho_{0}$ that the
the minimum  of $H_{c}^{r}$ is reached. Instead of this sharp 
cutoff, it is possible to introduce a smooth cutoff by adding to the
action an external source proportional to, for example,
$$
\int d \tau \sum_{\sigma} \left(1 -\rho(\sigma,\tau)\right)^{2}.
$$
However, in what follows, for the sake of simplicity, we will simply
fix the value of $\rho_{0}$ at some value less than one.

It seems that what we have done is to exchange one cutoff for
another: We have let the grid spacing $a$ go to zero, but instead,
 we have imposed an upper limit on the world sheet density of graphs
measured by $\rho$. There is, however, a big difference
between the two cutoffs.
 The grid in $\sigma$ distorts the world
sheet, and by eliminating it, the original  continuum
world sheet picture is recovered. In contrast, the cutoff imposed by
eq.(40) corresponds to a selection of the graphs;
we are putting an upper bound of $\kappa$ on the average
 number of solid lines and hence on the
number of propagators on the world sheet. The dynamics is still
represented by the same action (13); we have
simply chosen to study the set of of graphs subject to the
restriction (40). Later, we will discuss the delicate limit
$\kappa\rightarrow \infty$, $\rho_{0}\rightarrow 1$.

We end this section by a brief discussion of the ground state of
the model for $D=4$ (6 space-time dimensions). 
To find ground state for $D=4$, we set
$$
\frac{\partial  H_{c}^{r}}{\partial \lambda(\sigma)}= 0
$$
in eq.(25), which gives,
\be
\rho(\sigma)=\dbsm
\wss \,\frac{(\sigma_{2}-\sigma) (\sigma -\sigma_{1})}
{(\sigma_{2} -\sigma_{1})^{3}}.
\ee
Notice that in contrast to the case $D=2$, this equation does not fix
$\lambda_{0}$. Here $\lambda$ acts as a lagrange multiplier and sets
\be
 H_{c}^{r}=0.
\ee

We now evaluate the right hand side of the above  equation in the mean field
approximation: As before, the double sum is over only the hybrid
lines, where we set $\rho=\rho_{c}=\rho_{0}$. Solving 
for $L$, we have,
\be
L=4 \pi\,g_{r}^{2}\,\rho_{0}\,(1- \rho_{0})\,\sum_{n_{1}=1}^{\infty}
\sum_{n_{2}=1}^{\infty}\frac{n_{1}\,n_{2}}{(n_{1}+n_{2})^{3}}\,
(1-\rho_{0})^{n_{1}+n_{2}-2}.
\ee

In contrast to the case $D=2$, here $\lambda_{0}$ is arbitrary, but
 $L$ is determined in terms of  $\rho_{0}$ as before.

\vskip 9pt

\noindent{\bf 7. Quadratic Fluctuations Around The Classical Background}

\vskip 9pt

In this section, we will study the quantum fluctuations around the
classical background by expanding in powers of the field fluctuations
$\phi_{1}$ (eq.(14)).
 This expansion will be carried out only to second
order. Later, we will discuss how this expansion may be fitted into a
systematic perturbation series. In the interests of keeping this
paper to a resonable length, we will not try to do a complete second
order calculation; we will only study some chosen terms of interest.
To start with, $\rho$ and $\lambda$ will be frozen at their classical
given by eqs.(27) and (28), only the field $\phi$ will be expanded to
 second order in $\phi_{1}$. We should stress, however, there is no
obstacle to carrying out a complete second order calculation, except
for lack of interest.

It is convenient to set
$$
\phi_{1}= \phi_{1,r}+i\, \phi_{1,i},
$$
where $\phi_{1,r,i}$ are hermitian fields. The contribution
 to the action second order in $\phi_{1}$ is given by
\be
S^{(2)}= S_{k.e}- \int d\tau\,H^{(2)}(\tau)=S_{k.e}+ S_{p.e},
\ee
where,
\be
S_{k.e}=2 \sum_{\sigma} \int d\tau \int d\cue\,\phi_{1,i}\,
\partial_{\tau}\phi_{1,r},
\ee
 Since the action is quadratic in both $\phi_{1,i}$ or $\phi_{1,r}$,
 one can carry out the 
functional integral over one of these fields before writing down $H^{(2)}$.
 We choose to integrate over
 $\phi_{1,i}$, with the result,
\be
S_{k.e}\rightarrow  \sum_{\sigma} \int d\tau \int d\cue\,
\frac{\left(\partial_{\tau}\phi_{1,r}(\sigma,\tau,\cue)\right)^{2}}
{\lambda_{c}(\sigma)+ \frac{1}{2} \sum_{\sigma'} G(\sigma,\sigma')\,
\rho_{c}(\sigma')\,\cue^{2}},
\ee
and, somewhat schematically,
\bq
H^{(2)}&\rightarrow&\sum_{\sigma} \lambda_{c}(\sigma) \int d\cue\,
\phi_{1,r}^{2}(\sigma,\cue)+
\sum_{\sigma,\sigma'} G(\sigma,\sigma') \Bigg(\frac{1}{2}\,
\rho_{c}(\sigma')\,\int d\cue\,
\cue^{2}\,\phi_{1,r}^{2}(\sigma,\cue)\nonumber\\
& -& 2 \int d\cue \int d\cue' (\cue\cdot \cue')\,
(\phi_{0}\,\phi_{1,r})_{\sigma,\cue}\,(\phi_{0}\,\phi_{1,i})_{\sigma',
\cue'}\Bigg).
\eq
 As stated earlier, in this equation
 $\rho$ and $\lambda$ are fixed
at their classical values given by eqs.(27) and (28). We would like to
emphasize that the $\sigma$ sums in the above expressions are
over the hybrid lines which are spaced by $L$; the grid spacing $a$
has completely disappeared.

The above action, being quadratic in $\phi_{1,r}$, can be diagonalized.
 Again, in the interests of brevity, we will
not carry out a full analysis, but instead, describe the general
features of the spectrum, and work out in detail the sector of the
model of particular interest. One important feature is that limit
$a\rightarrow 0$ can be taken without causing any blowup or singularity
in the spectrum. This is in contrast to the earlier work [1,2], where,
in the limit $a\rightarrow 0$, part of the spectrum went to infinity
as $1/a^{2}$. This is due to different classical background we have
here: As we have pointed out earlier, the factor that is the source of  
possible singularity is $1/|\sigma-\sigma'|$ in $G(\sigma,\sigma')$.
But with the classical background we have, this factor never becomes
singular since, located on the hybrid lines, $\sigma$ and $\sigma'$ are
seperated by  at least a distance $L$, and $L$ stays fixed and finite
as  $a\rightarrow 0$.

So far, we have kept $\rho_{0}<1$. In the case $D=2$, this was done
with the help of a cutoff (eq.(40)). It is of considerable interest
to see what happens in the limit $\rho_{0}\rightarrow 1$. In this
 limit, which we will call the high density limit,
 $L\sim 1-\rho_{0}\rightarrow 0$ (eq.36), and from a
cursory examination of eq.(6), one expects the spectrum to blow up
as $1/L^{2}$. This in parallel with the  $1/a^{2}$ behaviour found
in [1,2]. Just as in that case, part of the spectrum
stays finite in this limit. This because the world sheet dynamics is
invariant under the translation
\be
\cue\rightarrow \cue+ {\bf r},
\ee 
where ${\bf r}$ is a constant vector. This invariance is broken
spontaneously by the classical solution, which is not translation
invariant. This situation is familiar from soliton and instanton physics;
as a consequence of Goldstone's theorem, a massless zero mode develops.
Protected by Goldstone's theorem, this mode remains massles also in
the limit $L\rightarrow 0$, and therefore there is no blowup in the
spectrum. We will call this sector of the model the light sector, and 
identify and quantize it by the collective coordinate method for
the case $D=2$. The modes whose energies go to infinity as 
$L\rightarrow 0$ will be called the heavy modes.

Consider the field configuration
\be
\phi_{1,r}\rightarrow \phi_{0}\left(\sigma,\cue+\bfv(\sigma,\tau)
\right)- \phi_{0}(\sigma,\cue),
\ee
where $\phi_{0}$ is the classical solution (17). What we have done is
to promote the constant vector ${\bf r}$ into the collective
coordinate $\bfv(\sigma,\tau)$. We now replace $\phi_{1,r}$ in
$S_{k.e}$ in eq.(46) by the above expression, expand to second order
in $\bfv$, and do the finite integral over $\cue$ explicitly. There
are couple of helpful simplifications which we note below:\\
a) Before applying the mean field approximation,
it is best to use the identities (18) and (20) in order to get rid of
$G$ and reduce quadratic terms in $W$ to linear ones. This simplifies
 the final expression considerably.\\
b) Since the classical solution in the mean field approximation is
$\tau$ independent, the only $\tau$ dependence is in $\bfv$. This
is why the final expression for $S_{k.e}$
 is quadratic in $\partial_{\tau}\bfv$:
\bq
S_{k.e}&\rightarrow&\frac{\pi\,g^{2}}{6} \int d\tau \sum_{\sigma}
\dbsm\frac{\wss}{(\sigma -\sigma_{1})\,(\sigma_{2} -\sigma_{1})\,
(\sigma_{2} -\sigma)\,\lambda_{0}^{3}}
\,\left(\partial_{\tau}
\bfv(\sigma,\tau)\right)^{2}\nonumber\\
&\rightarrow &\frac{\pi\,g^{2}\,\rho_{0}^{2}\,(1-\rho_{0})\,
f_{3}(\rho_{0})}{6\,\lambda_{0}^{3}\,L^{3}}\,\int d\tau
\sum_{\sigma} \left(\partial_{\tau} \bfv\right)^{2}.
\eq
Here $\lambda_{0}$ and $L$ are given by (36) and,
$$
f_{3}(\rho_{0})=\sum_{n_{1}=1}^{\infty} \sum_{n_{2}=1}^{\infty}
\frac{(1-\rho_{0})^{n_{1}+n_{2}- 2}}{n_{1}\,n_{2}\, (n_{1}+n_{2})}.
$$

Next, we make the replacement (49) in $H^{(2)}$ (eq.(47)). This is easily
evaluated by shifting $\cue$ integration by
$$
\cue\rightarrow \cue -\bfv
$$
 and then using 
$$
\int d\cue\, \phi_{0}^{\dagger}\,\phi_{0}= \rho.
$$
We have,
\bq
H^{(2)}&\rightarrow&
\frac{1}{2}\,\sum_{\sigma,\sigma'}
 G(\sigma,\sigma')\Bigg(\int d\cue
\,\rho(\sigma')\,\cue^{2}\,\left(\phi^{\dagger}_{0}
\phi_{0}\right)_{\sigma,\cue+\bfv(\sigma)}\nonumber\\
&-& \int d\cue \int d\cue' (\cue \cdot \cue')
\left(\phi^{\dagger}_{0}\phi_{0}\right)_{\sigma,\cue+\bfv(\sigma)}\,
\left(\phi^{\dagger}_{0}\phi_{0}\right)_{\sigma',\cue'+\bfv(\sigma')}
\Bigg)\nonumber\\
&\rightarrow& \frac{1}{2}\,\sum_{\sigma,\sigma'} G(\sigma,\sigma')
\,\rho(\sigma)\,\rho(\sigma') \left(\bfv^{2}(\sigma)-\bfv(\sigma)
\cdot \bfv(\sigma')\right).
\eq
Finally, applying the mean field approximation  to the
last line gives
\be
H^{(2)}\rightarrow \frac{1}{2} \rho_{0}^{2}\,\sum_{n \neq n'}
\frac{(1-\rho_{0})^{|n-n'| -1}}{|n-n'|\,L} \left( \bfv^{2}(
\sigma=n L) -\bfv(\sigma=n L)\cdot \bfv(\sigma'=n' L)\right).
\ee

The above expression can be diagonalized by defining
$$
\bfv(n L)= \frac{L}{2 \pi} \int_{-\pi/L}^{\pi/L} dk\, e^{- ik n L}\,
\tilde{v}(k),
$$
and rewriting it in terms of $\tilde{v}$:
\be
H^{(2)}\rightarrow\frac{\rho_{0}^{2}}{4 \pi\,(1-\rho_{0})}
\int_{-\pi/L}^{\pi/L} dk \tilde{v}(k)\cdot \tilde{v}(-k)\,
\ln\left(1+ \frac{2\,(1-\rho_{0})}{\rho_{0}^{2}}\left(1- \cos(k L)
\right)\right).
\ee
Up to an overall constant, the spectrum as a function of $k$ is
given by the function
$$
\ln\left(1+ \frac{2\,(1-\rho_{0})}{\rho_{0}^{2}}\left(1- \cos(k L)
\right)\right).
$$
This is a periodic function with the period $2 \pi/L$. This kind of spectrum
is to be expected in view of the formation of a periodic structure (crystal)
on the world sheet. We should point out that although we have treated
$k$ as a continuous variable, being conjugate to the compact variable 
$0\leq \sigma \leq p^{+}$, it is really a discrete variable quantized in
units of $2 \pi/p^{+}$. So the integral over $k$ should really be replaced by
a discrete sum. Since, however, we have tacitly assumed the ratio
$p^{+}/L$ to be large, the integral is a good approximation to the sum.

We now briefly discuss quadratic fluctuations in $\bfv$ for
 the case $D=4$. Again, starting with the field configuration (49),
we repeat the steps leading to eqs.(46) and (47).
 The result for $S_{p.e}$ is unchanged, again given by
eq.(53). $S_{k.e}$ can be computed again by applying eq.(46), leading to
\be
S_{k.e}\rightarrow \frac{\pi^{2}\,g_{0}^{2}\,\rho_{0}\,(1-\rho_{0})
f_{1}(\rho_{0})}{6\,\lambda_{0}^{2}\,L^{2}}\,\int d\tau \sum_{\sigma}
\left(\partial_{\tau} \bfv(\sigma,\tau)\right)^{2},
\ee
where $f_{1}$ is given by (37).

So far, $\lambda_{0}$ was arbitrary, but now, for the model to be
renormalizable, we must require the ratio
$$
g_{0}^{2}/\lambda_{0}^{2}
$$
to be cutoff independent, so that
\be
\lambda_{0}^{2}\rightarrow \frac{1}{\ln\left(\Lambda^{2}/\mu^{2}\right)}
\ee
as $\Lambda\rightarrow \infty$.

\vskip 9pt

\noindent{\bf 8. Dense Graphs On The World Sheet}

\vskip 9pt

We start with $D=2$, and take the limit of dense graphs:
  $\rho_{0}\rightarrow 1$, $L\rightarrow 0$ in both $S_{k.e}$ (eq.(50))
and in
$$
S_{p.e}=\int d\tau\,H^{(2)}(\tau).
$$
In this limit, two simplifications occur:\\
a) As $\rho_{0}\rightarrow 1$, the action becomes local. By a local action
we mean an action which correlates two $\bfv$'s seperated by at most
by a distance $L$. Notice that $S_{k.e}$ is already local, but $H^{(2)}$
contains terms of the form
$$
\bfv(n L)\cdot \bfv((n+n') L)
$$
with $n'>1$, which are non-local. However, in the limit
 $\rho_{0}\rightarrow 1$, all of these non-local terms are suppressed
by factors of $1- \rho_{0}$.\\
b) As $L\rightarrow 0$, $\sigma$ becomes continuous. The sums over
$\sigma$ turn into integrals, and one can expand in powers of $L$:
$$
\bfv((n+1) L)- \bfv(n L)\rightarrow L\,\partial_{\sigma}\bfv(\sigma).
$$

We start with $S_{k.e}$. We let $\rho_{0}\rightarrow 1$, or,
equivalently, $L\rightarrow 0$, use eq.(36) for
$\lambda_{0}$, and convert the sum
over $\sigma$ into an integral:
\be
S_{k.e}\rightarrow \frac{C}{2\,\mu^{4}}\,\int d\tau \int d\sigma(
\partial_{\tau}\bfv)^{2},
\ee
where C is a numerical constant:
$$
C=\frac{\exp(2/3)}{48 \pi^{2}}.
$$
The important point is that,  $S_{k.e}$ is independent
of $L$, so that a finite limit is reached as $L\rightarrow 0$.
 We also note that the result is independent of $g$.

Next, we consider eq.(53). Expanding to  leading
order in $L$ by
$$
1 -\cos(k L)\rightarrow \frac{1}{2}\, k^{2}\,L^{2},
$$
and letting $\rho_{0}\rightarrow 1$ gives
\be
S_{p.e}\rightarrow -\frac{L^{2}}{4 \pi}\,\int d\tau \int dk
\,k^{2}\,\tilde{v}(\tau,k)\cdot \tilde{v}(\tau, -k) =
- \frac{1}{2}\,\int d\tau \int d\sigma \,\left(\partial_{\sigma}
\bfv(\tau,\sigma)\right)^{2}.
\ee
The limit is again finite and $g$ independent. Finally, the limit
of the full second order action is given by
\be
S^{(2)}\rightarrow \int d\tau \int_{0}^{p^{+}} d\sigma
\left(\frac{C}{2\,\mu^{4}}\,\left(\partial_{\tau}\bfv\right)^{2}
-\frac{1}{2}\,\left(\partial_{\sigma}\bfv\right)^{2}\right).
\ee

This the action for a transverse string in the lightcone
coordinates. The slope is given by
\be
\alpha'= \left(4 \pi^{2}/C\right)^{1/2}\,\mu^{2},
\ee
and depends only on $\mu$.
 It is important to notice that the string picture is an
 approximate one. In reality, $1-\rho_{0}$ is small but not zero,
and there is a heavy sector with masses proportional to
$$
1/L^{2}\simeq 1/(1-\rho_{0})^{2}.
$$
The higher string excitations with masses comparable to the masses
of the states in the heavy sector  mix with these states,
 and the string picture breaks down. It is plausible to
  identify the heavy sector with the original field theory spectrum:
Since the model at $D=2$ is trivially asymptotically free, a weakly coupled
field $\phi^{3}$ theory is expected to be valid at high energies.
On the other hand, at low energies, the picture developed here
suggests that low lying  bound states of the model form a string.
 So we have a
hybrid picture combining field theory and string theory: At low energies,
the string picture is the relevant one, and at high energies, field theory
takes over.

It is of interest to notice that in the dense graph limit, the
combination
\be
\gamma=g^{2}\,(1-\rho_{0})
\ee
acts as an effective coupling constant. For example, the classical
action (38) is proportional to $1/\gamma^{2}$, the lowest order
quantum corrections (58) are independent of $\gamma$. This is the
behaviour expected from a weak coupling expansion in $\gamma$ and it can
be traced back to the structure of the interaction term  in
eq.(1)). From this
perspective, the heavy sector, with masses inversely proportional 
to $\gamma$, can be thought of as a sector of solitons. In the dense
graph limit, it may be possible to do a systematic expansion in
$\gamma$, without any appeal to the mean field approximation.
 We hope to further develop this idea in the future.

Finally, we comment briefly on the dense graph limit at $D=4$.
This limit is more problematic in this case, since $\lambda_{0}$
is undetermined. We note that, as $\rho_{0}\rightarrow 1$,
$L\sim 1-\rho_{0}$ (eq.(36)), and in order to have a sensible string
picture with a finite slope, we have to require $S_{k.e}$
(eq.(54)) to remain finite. Therefore, in this limit,
in addition to eq.(55), the
following condition has to be imposed on $\lambda_{0}$:
$$
\lambda_{0}\sim 1/(1-\rho_{0}).
$$
From the string perspective, this is a natural requirement:
We are demanding that the string slope remain finite
 in the dense graph limit. However, it would be very desirable
to confirm this by means of of an alternative treatment based
on the original field theory model.

\vskip 9pt

\noindent{\bf 9. Conclusions}

\vskip 9pt

The main contribution of the present article is the correct handling of
the two divergences that plagued the previous work [1,2] on the world
sheet field theory [5] for the $\phi^{3}$ interaction. The ultraviolet
divergence is eliminated by the mass and coupling constant terms
already present in the model, without any ad hoc modifications, which
was an unsatisfactory feature of [1]. The other problem is a spurious
infrared divergence, which requires the discretization of the world sheet
coordinate $\sigma$. Here we are able
 to take the limit grid spacing $a\rightarrow 0$ smoothly, without
encountering any blow up in the spectrum. This is achieved by 
choosing a classical background different from the one chosen in [1,2].
The new background consists of equally spaced parallel lines that form
a one dimensional crystal. Possible divergences are avoided by
keeping the spacing of lines fixed as $a\rightarrow 0$.

We feel that, except for manifest Lorentz invariance\footnote{See [10]
for an investigation of renormalization and Lorentz invariance
in the light cone formulation.}
 all the major technical problems associated with the $\phi^{3}$
model have been resolved, at least within the context of the mean
field approximation. It is time to apply the techniques
developed for $\phi^{3}$ to more physical models, such as gauge
theories\footnote{For some initial steps taken towards more
physiacal theories, see [11, 12].}
 An intermediate step would be to introduce, in addition to
$\phi^{3}$, a $\phi^{4}$ interaction. This is a more physical model,
and to some extent, mimics a gauge theory. We hope to investigate this
 possibility in the near future.

\vskip 9pt

\noindent{\bf Acknowledgement}

\vskip 9pt

This work was supported in part by the director, Office of Science,
Office of High Energy Physics of the U.S. Department of Energy under Contract
DE-AC02--05CH11231.

\vskip 9pt

\noindent{\bf References}

\vskip 9pt

\begin{enumerate}

\item K.Bardakci, JHEP {\bf 1003} (2010) 107, arXiv:0912.1304.
\item K.Bardakci, JHEP {\bf 0903} (2009) 088, arXiv:0901.0949.
\item K.Bardakci and C.B.Thorn, Nucl.Phys. {\bf B 626} (2002)
287, hep-th/0110301.  
\item G.'t Hooft, Nucl.Phys. {\bf B 72} (1974) 461.
\item K.Bardakci, JHEP {\bf 0810} (2008) 056, arXiv:0808.2959.
\item H.P.Nielsen and P.Olesen, Phys.Lett. {\bf B 32} (1970) 203.
\item B.Sakita and M.A.Virasoro, Phys.Rev.Lett. {\bf 24} (1970) 1146.
\item A.Casher, Phys.Rev. {\bf D 14} (1976) 452.
\item R.Giles and C.B.Thorn, Phys.Rev. {\bf D 16} (1977) 366.
\item C.B.Thorn, Nucl.Phys. {\bf B 699} 427, hep-th/0405018,
D.Chakrabarti, J.Qiu and C.B.Thorn, Phys.Rev. {\bf D 74} (2006)
045018, hep-th/0602026.
\item C.B.Thorn, Nucl.Phys. {\bf B 637} (2002) 272, hep-th/0203167,
S.Gudmundsson, C.B.Thorn and T.A.Tran, Nucl.Phys. {\bf B 649} 92003)
3-38, hep-th/0209102.
\item C.B.Thorn and T.A.Tran, Nucl.Phys. {\bf B677} (2004) 289,
hep-th/0307203.

\end{enumerate}

\end{document}